# Quantum-Circuit Residuals for Receptor Functional Potency Prediction: Predictive Gains and Quadratic Structure

Chenyan Jia,[†] Yaping Ma,[‡] Yuen Wang,[‡] and Cong Guo[*,†]

†*Shenzhen SpinQ Technology Co., Ltd., Shenzhen 518048, China*
‡*Shenzhen DIVBIO Pharmaceutical Co., Ltd., Shenzhen 518057, China*

E-mail: cguo@spinq.com

**Abstract**

The predictive performance of neural networks for receptor functional potency modeling depends on both molecular representations and prediction-head design. Using GLP-1R cAMP-related $pEC_{50}$ prediction as the primary task, we jointly train a classical molecular encoder with a linear base head and a shallow quantum-circuit residual head that share the learned molecular representation. We evaluate the models using random and structure-group splits, include tuned Chemprop as a domain reference, and extend the evaluation to ADRB2 and MC4R. Under the GLP-1R structure-group split, the quantum-residual models reduce mean absolute error relative to the corresponding base models from $0.677 \pm 0.010$ to $0.639 \pm 0.016$ with a multilayer perceptron (MLP) encoder and from $0.680 \pm 0.009$ to $0.641 \pm 0.027$ with a graph neural network (GNN) encoder (mean $\pm$ sample SD). In structural control experiments, a parameter-count-matched real-orthogonal control achieves similar predictive performance, suggesting that some quantum-residual gains may be associated with structured quadratic feature interactions. By combining receptor functional potency prediction with structural analysis of the quantum residual, this study provides empirical evidence to inform compact residual-head design and highlights the potential of quantum-circuit parameterizations for this task.

# 1 Introduction

The glucagon-like peptide-1 receptor (GLP-1R) is an important therapeutic target for metabolic diseases. The therapeutic benefits of GLP-1R agonists in glycemic control and weight management have motivated continued efforts to develop peptide and nonpeptide small-molecule agonists.[1–4] The cyclic adenosine monophosphate (cAMP) response provides a functional readout of receptor activation for these structurally diverse agonists. Potency in these assays is quantified by the half-maximal effective concentration ($EC_{50}$), the concentration required to elicit half of the maximal response under the specified assay conditions.[5] Predicting cAMP-related $pEC_{50}$, defined as the negative base-10 logarithm of $EC_{50}$ expressed in molar units, from molecular structure could therefore support the prioritization of compounds for experimental evaluation.

Neural approaches to molecular property prediction use molecular fingerprints, representations learned through message passing over molecular graphs, or combinations of the two.[6] Chemprop provides an established framework for jointly learning molecular representations and property predictions.[7,8] Beyond the molecular representation, predictive performance also depends on the function that maps latent features to the target property. Adding a residual head to a linear base head provides a framework for examining this mapping. With both heads operating on a shared representation and trained jointly with the encoder, different residual forms and parameterization constraints can be evaluated within the same predictive architecture.

Parameterized quantum circuits provide a class of predictive modules whose functional form is determined by data encoding, trainable circuit transformations, and observable readout.[9,10] Hybrid quantum–classical models have been investigated for predicting drug toxicity, molecular physicochemical properties, and protein–ligand binding.[11–13] These approaches combine classical molecular representations with quantum regressors or fusion modules, including models based on multilayer perceptrons (MLPs) and graph neural networks (GNNs).[12,13] Related work on peptide–HLA binding prediction has examined quantum modules across different training-set sizes.[14] Evaluating these models against classical alternatives and analyzing their functional structure can help clarify how model design relates to predictive performance.[15,16] We therefore examine whether a jointly trained quantum-circuit residual improves receptor functional potency prediction and how structured classical controls inform the interpretation of its predictive gains.

The main contributions of this study are as follows:

(1) We develop a hybrid quantum–classical model for receptor functional potency prediction that jointly trains a molecular encoder, a linear base head, and a compact quantum-circuit residual head, with both heads sharing the learned molecular representation.

(2) We evaluate predictive performance and residual gains on GLP-1R cAMP-related $pEC_{50}$ prediction under random and structure-group splits, with separate evaluations on ADRB2 and MC4R to assess their dependence on receptor task, encoder, and split scheme.

(3) We characterize the quantum residual as a constrained quadratic form on the normalized latent representation and compare it with diagonal, low-rank, and parameter-count-matched real-orthogonal controls to investigate the relationship between residual structure and predictive gains.

# 2 Methods

## 2.1 Data Curation and Prediction Tasks

Activity records for the primary GLP-1R task and the additional ADRB2 and MC4R tasks were obtained from ChEMBL 37.[17] Eligible records were associated with the corresponding human single-protein targets, had assay descriptions referring to cAMP, and reported positive, finite $EC_{50}$ values with an exact relation ("=") and units convertible to molar concentration. Records flagged for data-validity issues or potential duplication, or explicitly involving antagonism, allosteric modulation, or receptor variants, were excluded.

Molecular structures were standardized using RDKit, including removal of salt counterions and charge neutralization where applicable, while preserving stereochemical information.[18] After conversion of $EC_{50}$ values to $pEC_{50}$, records sharing the full InChIKey of the standardized structure were aggregated within each target, and their median $pEC_{50}$ was used as the prediction label.[19] The resulting labels summarize the available measurements but remain subject to variability across assay conditions.[20] Record selection and variability among repeated measurements are described in Supporting Information Section S1.

For descriptive and stratified analyses, compounds were assigned to operational groups based on the number of amide bonds in their standardized structures: no more than two for the small-molecule group and three or more for the peptide group. Models were developed separately for each receptor, with both compound groups included in the corresponding dataset. Dataset composition is summarized in Table 1.

## 2.2 Data Splitting and Evaluation Design

Each receptor dataset was evaluated using five-fold random and structure-group splits.[6,21] Random splits were stratified by $pEC_{50}$. For structure-group splits, molecular similarity was calculated using chirality-aware Morgan fingerprints (radius 2, 2048 bits).[18,22] Molecules were linked if their Tanimoto similarity was at least 0.90 or they shared the same InChIKey connectivity block. The resulting connected components, including molecules linked indirectly through other members, were assigned as intact groups to training, validation, and test sets while balancing sample counts and $pEC_{50}$ distributions. Further details are provided in Supporting Information Section S2.

Each fold served once as the test set, while the remaining samples were divided into training and validation sets, giving approximate overall proportions of 70:10:20 for training, validation, and testing. Each compound received one out-of-fold test prediction per training seed. For each receptor and split scheme, all models used the same fixed data partitions (split seed 42). The models developed in this study were trained with five random seeds (42–46), with validation MAE used for model selection. Out-of-fold predictions were pooled across the five folds before calculating the evaluation metrics, as described in Section 2.6.

## 2.3 Classical–Quantum Residual Architecture and Joint Optimization

The model comprises a molecular encoder, a linear base head, and an additive residual head, all trained jointly (Figure 1). The encoder $f_\phi$, with parameters $\phi$, maps molecular input $x$ to a 32-dimensional real-valued representation $h = f_\phi(x)$. The base head operates on $h$, while the residual head receives its L2-normalized form, $\widehat{h}$. The final prediction is

$$\widehat{t} = t_0 + \Delta(\widehat{h}). \tag{1}$$

$$t_0 = w_0^\mathsf{T} h + b_0. \tag{2}$$

Here, $\widehat{t}$ is the predicted standardized $pEC_{50}$, $t_0$ is the base prediction, and $\Delta$ is the residual function. The vector $w_0$ and scalar $b_0$ are the base-head weights and bias, respectively, and $\mathsf{T}$ denotes transpose. Within each fold, labels are standardized using the training-set mean and population standard deviation. Predictions are transformed back to the original $pEC_{50}$ scale before evaluation.

The MLP encoder maps chirality-aware 256-bit Morgan fingerprints (radius 2) to the 32-dimensional representation through two hidden layers with GELU activations and a tanh output layer.[23] The GNN encoder combines a graph representation obtained by directed bond message passing and mean pooling[6] with the same molecular fingerprint by concatenation, then projects the combined features into the same 32-dimensional latent space. Supplementary evaluations include a SMILES-based Transformer encoder with multihead self-attention[24] and a 32-dimensional output. Architectural details are provided in Supporting Information Section S3.1, and Transformer results are reported in Table S5 and discussed in Section S6.5 of the Supporting Information.

The quantum residual head uses amplitude encoding to represent the normalized vector $\widehat{h}$ as a five-qubit input state:[25]

$$|\psi(\widehat{h})\rangle = \sum_{j=0}^{31} \widehat{h}_j |j\rangle. \tag{3}$$

Here, $|\cdot\rangle$ denotes a quantum state vector, and $|\psi(\widehat{h})\rangle$ is the encoded input state. The index $j$ runs over the $2^5 = 32$ computational basis states $|j\rangle$, labeled by five-bit binary strings. Each real-valued component $\widehat{h}_j$ specifies the amplitude of the corresponding basis state.

The encoded state passes through two parameterized circuit layers. Each layer applies rotations $R_X$, $R_Y$, and $R_Z$ about the $x$, $y$, and $z$ axes, respectively, to each qubit, followed by CNOT gates arranged in a ring. A CNOT (controlled-NOT) gate flips the target qubit conditional on the control qubit being in state $|1\rangle$. The circuit implements a unitary transformation $U_\theta$, parameterized by the trainable rotation angles $\theta$. The quantum residual is a

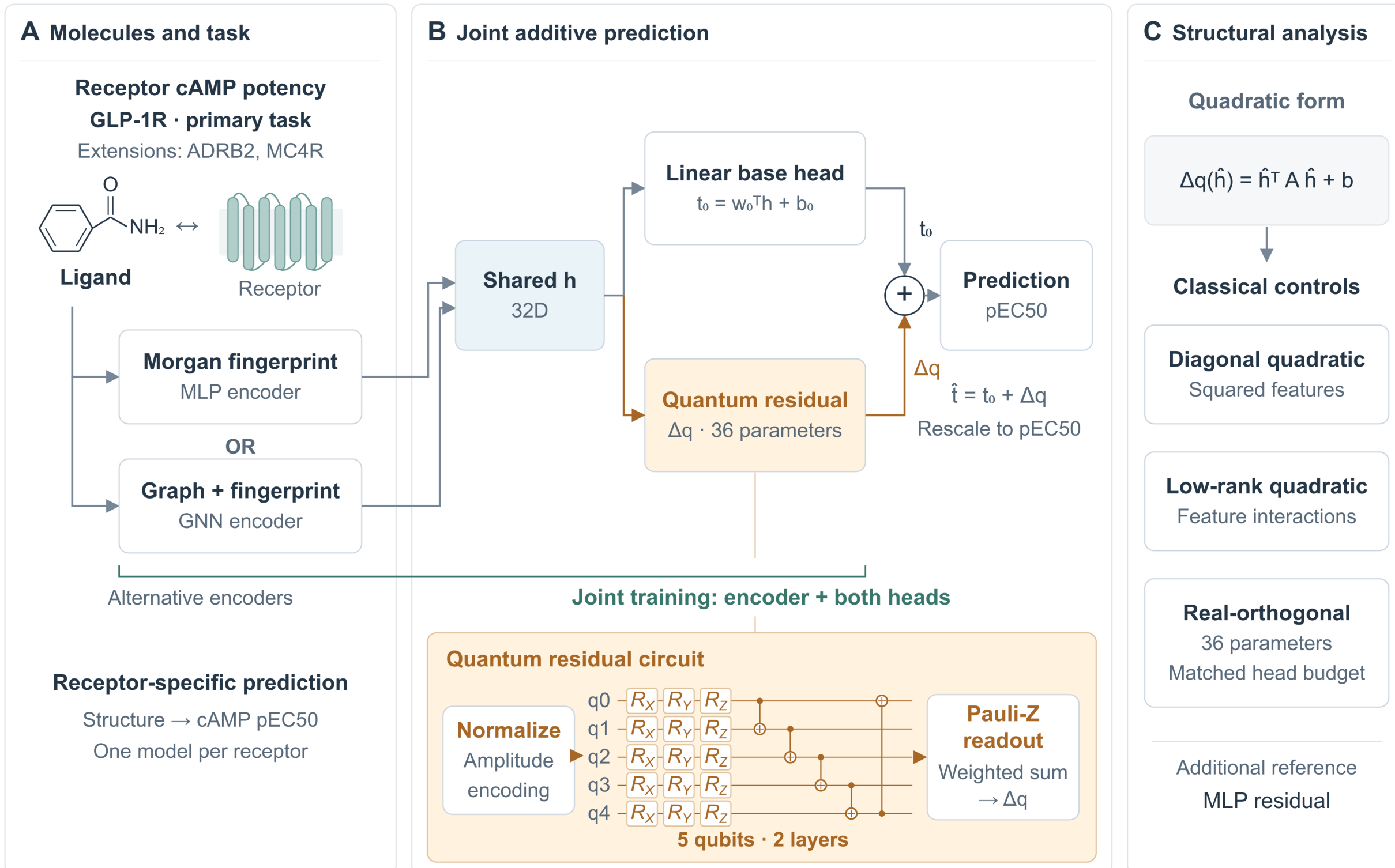

Figure 1: Quantum-circuit residual framework and structured controls. (A) Ligand structures are used to predict receptor cAMP-related $pEC_{50}$, with each receptor modeled separately. The MLP takes Morgan fingerprints, whereas the GNN combines molecular graphs and fingerprints. (B) The shared representation feeds the linear base head and quantum residual head; their outputs are summed and transformed back to the $pEC_{50}$ scale. All three components are trained jointly end to end. The light-orange module shows amplitude encoding, a two-layer five-qubit circuit, and single-qubit Pauli-$Z$ expectation-value readout. (C) Quadratic structure of the quantum residual and classical controls. The coefficient matrix $A$ is defined in the text, and each variant is trained jointly.

weighted sum of the five single-qubit Pauli-$Z$ expectation values plus a bias:

$$\Delta_{\mathrm{q}}(\widehat{h}) = \sum_{k=0}^{4} a_k \langle \psi(\widehat{h}) | U_\theta^\dagger Z_k U_\theta | \psi(\widehat{h}) \rangle + b. \tag{4}$$

The subscript q denotes the quantum residual, $k = 0, \ldots, 4$ indexes the qubits, and $a_k$ and $b$ are the trainable readout weights and bias, respectively. The operator $Z_k$ acts as Pauli-$Z$ on qubit $k$ and as the identity on the other qubits. Its expectation value after the circuit equals the probability of measuring qubit $k$ in state $|0\rangle$ minus that of measuring it in state $|1\rangle$. The symbol $\dagger$ denotes conjugate transpose, and $\langle\psi|$ is the conjugate transpose of $|\psi\rangle$. The quantum residual head contains 36 trainable parameters: 30 rotation angles, five readout weights, and one bias.

Quantum expectation values are computed using differentiable state-vector simulation implemented in PyTorch.[26] The encoder, base head, and residual head are jointly optimized by minimizing the Smooth L1 loss between the final prediction and the standardized target. Training uses AdamW[27] with a cosine learning-rate schedule.[28] Early stopping and model selection are based on validation MAE. Training settings are provided in Supporting Information Section S4.1.

For the real-valued normalized representation $\widehat{h}$, the quantum residual can be written as a constrained quadratic form:

$$\Delta_{\mathrm{q}}(\widehat{h}) = \widehat{h}^{\mathsf{T}} A \widehat{h} + b. \tag{5}$$

$$A = \mathrm{Re}(U_\theta^\dagger D U_\theta). \tag{6}$$

$$D = \sum_{k=0}^{4} a_k Z_k. \tag{7}$$

Here, $D$ is the real diagonal readout matrix, and Re denotes the real part. The $32 \times 32$ real symmetric matrix $A$ is parameterized by the circuit rotation angles and readout weights, with its form constrained by the circuit architecture and readout operators. This expression describes the residual as a function of $\widehat{h}$, while the encoder is optimized jointly with the prediction heads. The gate sequence and derivation are provided in Supporting Information Section S3.2.

## 2.4 Classical Residual Controls

The base model used the same encoder architecture and a linear prediction head without a residual branch. A compact MLP residual head was included as a classical nonlinear control. Additional controls comprised diagonal, low-rank, and real-orthogonal quadratic residual heads (Table 2). These use weighted squared features, quadratic combinations of two projected features, and orthogonal feature mixing followed by a quadratic readout, respectively. The real-orthogonal head matches the quantum residual head in trainable parameter count. Throughout, "MLP residual" denotes the additive prediction head rather than the molecular encoder.

The real-orthogonal control applies two layers of trainable rotations and fixed coordinate permutations to the normalized representation, preserving its norm. The transformed features are squared and summed using the same five fixed sign patterns as the quantum Pauli-$Z$ readout, followed by a learned linear combination and bias. This gives 36 trainable parameters—30 rotation angles, five readout weights, and one bias—matching the quantum residual head. The formulation is provided in Supporting Information Section S3.3.

Each residual model jointly trained its own encoder, base head, and residual head. Comparisons within each encoder architecture used identical data splits, training seeds, and training settings. Diagonal, low-rank, and real-orthogonal controls were evaluated with MLP and GNN encoders. Supplementary evaluations using the Transformer encoder included the base, MLP-residual, and quantum-residual models; their results are reported in Supporting Information Table S5 and discussed in Section S6.5.

Table 2: Residual-head structures, trainable parameter counts, and evaluation scope. All residual heads take the L2-normalized 32-dimensional representation and produce a scalar added to the base prediction. Counts include biases but exclude the encoder and base head; zero denotes the absence of a residual branch. All listed model–encoder combinations were evaluated on three targets, two split schemes, and five training seeds.

| Model/control | Residual structure | Param. | Encoders |
| --- | --- | --- | --- |
| Base | Linear base prediction; no residual branch | 0 | MLP/GNN/ Transformer |
| MLP residual | Two-unit tanh hidden layer followed by a linear output | 69 | MLP/GNN/ Transformer |
| Quantum residual | Five-qubit amplitude encoding, two circuit layers, and a linear readout of five single-qubit Pauli-$Z$ expectation values | 36 | MLP/GNN/ Transformer |
| Diagonal quadratic | Weighted sum of squared input features | 33 | MLP/GNN |
| Low-rank quadratic | Weighted sum of squares and the cross product of two learned linear projections | 68 | MLP/GNN |
| Real-orthogonal quadratic | Two layers of real rotations and fixed permutations, followed by a quadratic readout using five fixed sign patterns | 36 | MLP/GNN |

## 2.5 Domain Reference and Hyperparameter Selection

A directed message-passing neural network implemented in Chemprop 2.3.1 served as the domain reference.[6–8] We evaluated 12 configurations varying message-passing depth, hidden dimension, and dropout, selecting the configuration with the lowest mean validation MAE for each target and split scheme. Tuning details and selected configurations are provided in Supporting Information Section S4 and Table S4a.

Hyperparameter selection and out-of-fold evaluation shared the same data partition, introducing dependence between selection and evaluation. Chemprop results are therefore interpreted as a domain reference, while residual gains are quantified through paired comparisons within each encoder architecture. Additional Chemprop results for the default configuration and the adjusted training protocol are provided in Supporting Information Table S5b.

## 2.6 Evaluation Metrics and Statistical Analysis

Mean absolute error (MAE) was the primary metric, supplemented by root-mean-square error (RMSE) and the coefficient of determination ($R^2$).[29] Metrics were calculated from pooled out-of-fold predictions for each training seed, with MAE and RMSE expressed on the original $pEC_{50}$ scale. Results are reported as the mean ± sample standard deviation across seeds, reflecting training variability under fixed data splits.

Paired MAE differences are reported as the mean ± sample standard deviation. Differences were defined as base-model MAE minus quantum-residual MAE in Tables 3 and 4, and control MAE minus quantum-residual MAE in Figure 3 and Supporting Information Table S6a. Positive values indicate improvement relative to the corresponding base or control model.

For comparisons among the models developed here, 95% percentile intervals for MAE differences were estimated from 10,000 paired bootstrap resamples of structural connected components.[30,31] These intervals reflect evaluation-sample uncertainty with predictions held fixed. Calculation details and complete metrics are provided in Supporting Information Section S6 and Table S5; paired results and intervals are available in the "Paired Comparisons" worksheet of Supplementary Data S1.

# 3 Results

## 3.1 Dataset Composition

The final datasets contained 1052 GLP-1R, 829 ADRB2, and 767 MC4R compounds (Table 1). Under the amide-count-based grouping, GLP-1R and MC4R contained comparable numbers of compounds in the small-molecule and peptide groups, whereas ADRB2 was predominantly composed of the small-molecule group. Prediction errors stratified by compound group are reported in Supporting Information Section S6.4 and Table S6b.

Table 1: Dataset composition for the three receptor tasks. Operational groups comprise compounds with $\leq 2$ amide bonds (small-molecule group) or $\geq 3$ amide bonds (peptide group).

| Target | Original records | Eligible records | Aggregated compounds | Small-molecule group | Peptide group |
|---|---|---|---|---|---|
| GLP-1R | 2365 | 1147 | 1052 | 472 | 580 |
| ADRB2 | 1897 | 909 | 829 | 711 | 118 |
| MC4R | 2081 | 886 | 767 | 379 | 388 |

## 3.2 Quantum Residuals Improve GLP-1R Functional Potency Prediction

For GLP-1R, quantum-residual models achieved lower mean MAE than the corresponding base and MLP-residual models with both MLP and GNN encoders under random and structure-group splits (Figure 2A,B and Table 3). MAE was lower than that of the corresponding base model for every training seed in all four encoder–split combinations (Supplementary Data S1, "Paired Comparisons" worksheet).

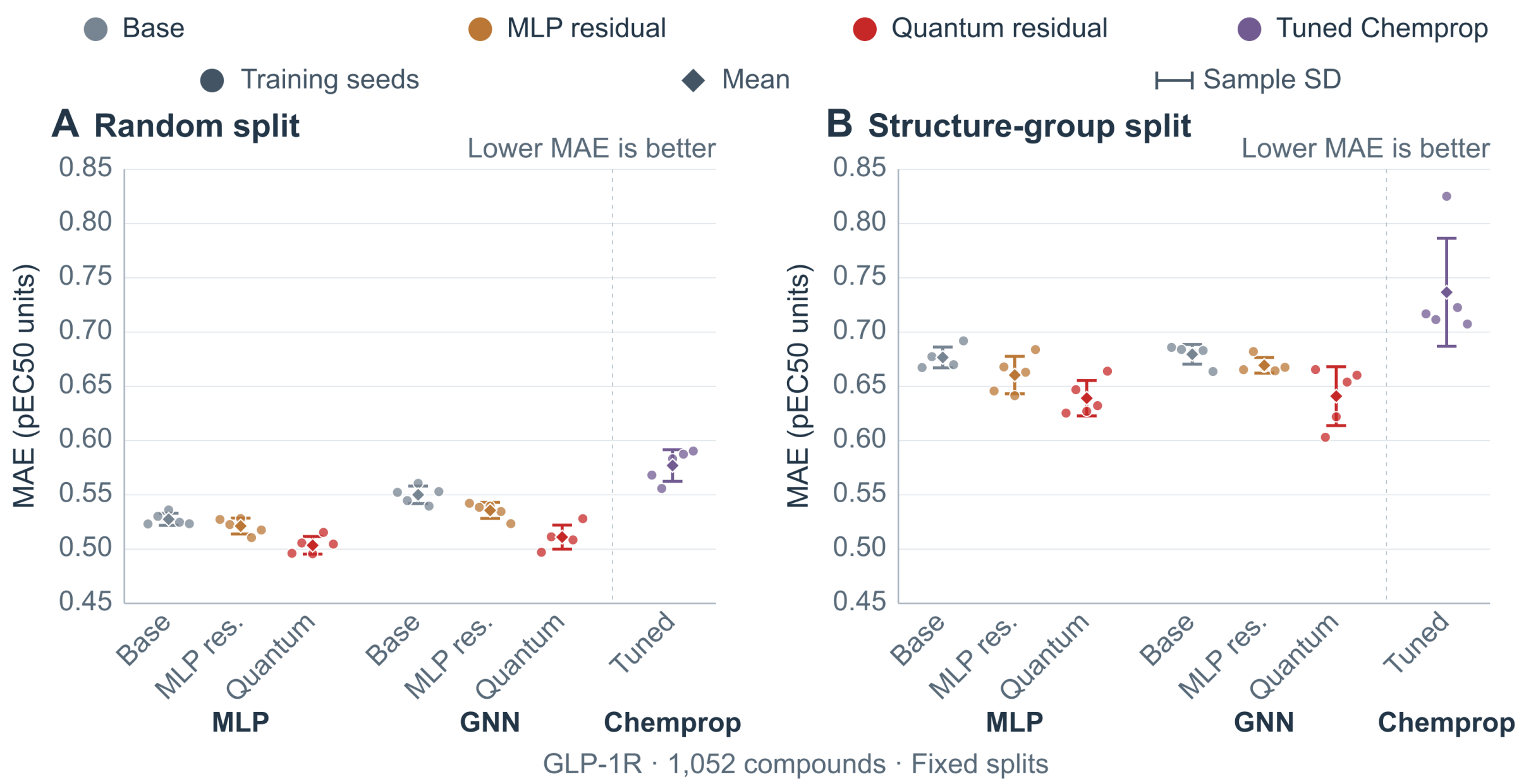

Figure 2: GLP-1R functional potency prediction. (A,B) Out-of-fold MAE under random and structure-group splits, respectively; lower values indicate better performance. Circles denote individual-seed results, diamonds denote means, and error bars show sample standard deviations. MLP/GNN denote encoders, and MLP res. denotes the MLP residual head.

Table 3: GLP-1R prediction results and paired improvements from quantum residuals.

| Split | Encoder/ reference | Base MAE | MLP residual MAE | Quantum residual MAE | Tuned Chemprop MAE | Paired improvement |
|---|---|---|---|---|---|---|
| Random | MLP | $0.528 \pm 0.006$ | $0.521 \pm 0.007$ | $0.504 \pm 0.008$ | — | $0.024 \pm 0.011$ |
| | GNN | $0.550 \pm 0.008$ | $0.536 \pm 0.007$ | $0.511 \pm 0.011$ | — | $0.039 \pm 0.013$ |
| | Chemprop | — | — | — | $0.577 \pm 0.015$ | — |
| Structure-group | MLP | $0.677 \pm 0.010$ | $0.660 \pm 0.017$ | $0.639 \pm 0.016$ | — | $0.038 \pm 0.009$ |
| | GNN | $0.680 \pm 0.009$ | $0.669 \pm 0.007$ | $0.641 \pm 0.027$ | — | $0.039 \pm 0.031$ |
| | Chemprop | — | — | — | $0.737 \pm 0.050$ | — |

Values are means ± sample standard deviations across training seeds, in $pEC_{50}$ units. Paired improvement is base-model MAE minus quantum-residual MAE; positive values indicate improvement over the base model. — indicates not applicable. Complete metrics and paired comparisons are provided in Supporting Information Tables S5 and S6a.

All models had higher mean MAE under the structure-group split than under the random split. Quantum-residual models nevertheless retained improvements over their corresponding base models, with paired MAE reductions of $0.038 \pm 0.009$ $pEC_{50}$ units for MLP and $0.039 \pm 0.031$ for GNN (Table 3).

MLP-residual models also reduced mean MAE relative to the base models in all four settings, but the reductions were smaller than those achieved by quantum-residual models (Figure 2A,B). The quantum residual head used 36 trainable parameters, compared with 69 for the MLP residual head, and achieved lower mean MAE across these settings.

The MLP and GNN base models already achieved lower mean MAE than tuned Chemprop under both split schemes (Table 3). Adding quantum residuals further reduced MAE for both encoder architectures, demonstrating incremental predictive gains over base models that were already competitive with an established molecular prediction framework.

## 3.3 Structural Controls for Quantum Residuals

Comparisons with diagonal, low-rank, and real-orthogonal quadratic controls examined how the predictive gains related to residual structure (Table 2). For GLP-1R, both quantum-residual and real-orthogonal models achieved lower mean MAE than the diagonal and low-rank quadratic controls across both encoders and split schemes (Figure 3 and Supporting Information Table S5).

Under the random split, quantum-residual and real-orthogonal models had similar mean MAE, with paired differences of $0.005 \pm 0.012$ $pEC_{50}$ units for MLP and $-0.003 \pm 0.011$ for GNN (Figure 3A,C). Under the structure-group split, quantum-residual models achieved lower mean MAE with both encoders; the corresponding differences were $0.020 \pm 0.011$ and $0.008 \pm 0.008$ $pEC_{50}$ units, respectively (Figure 3B,D). Complete metrics and paired comparisons are provided in Supporting Information Tables S5 and S6a.

The performance of the parameter-count-matched real-orthogonal control suggests that structured quadratic feature interactions may contribute to the predictive gains of quantum residuals.

## 3.4 Predictive Performance and Structural Controls on ADRB2 and MC4R

Quantum-residual models achieved lower mean MAE than their corresponding base models on both ADRB2 and MC4R across both encoders and split schemes (Figure 4 and Table 4). For ADRB2, reductions were larger under the random split. Under the structure-group split, mean MAE was close to the base-model value with the MLP encoder and slightly lower with GNN (Figure 4A).

For MC4R, quantum-residual models reduced mean MAE relative to their base models with both encoders under both split schemes, although the reductions were smaller under the structure-group split (Figure 4B). MLP-residual models also reduced mean MAE in three of the four encoder–split combinations; the exception was the GNN encoder under the structure-group split, where mean MAE exceeded that of the base model.

Real-orthogonal residuals achieved mean MAE close to that of quantum-residual models across both receptors, encoders, and split schemes (Table 4). Their relative rankings varied with the setting: under the MC4R structure-group split, quantum residuals had slightly lower mean MAE with MLP, whereas real-orthogonal residuals had slightly lower mean MAE with GNN. These results are consistent with the GLP-1R findings and suggest that structured quadratic feature interactions may contribute to predictive gains across receptor tasks.

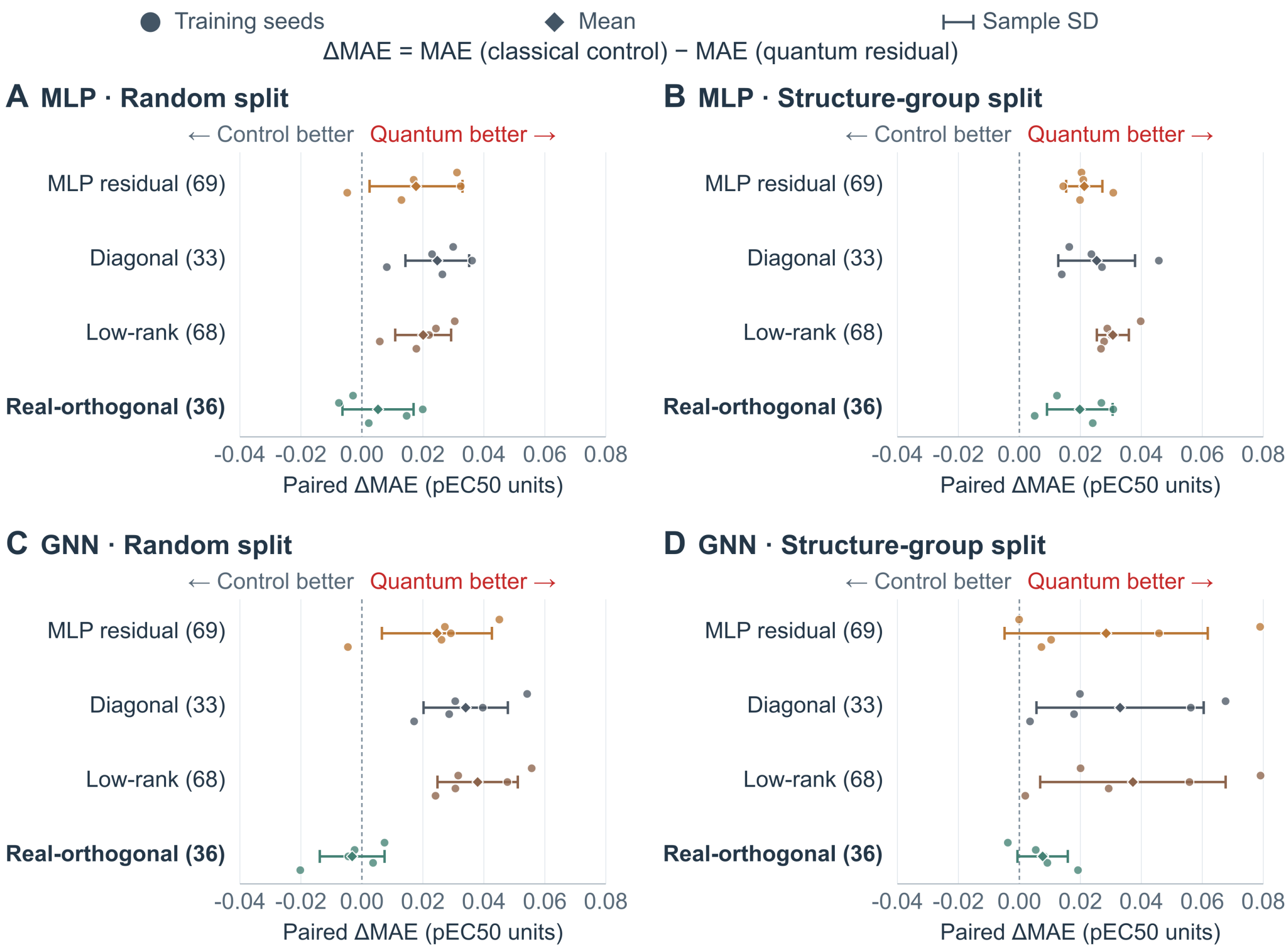

Figure 3: Paired MAE comparisons of classical residual controls and quantum-residual models on GLP-1R. (A,B) MLP encoder under random and structure-group splits, respectively; (C,D) GNN encoder under the corresponding splits. Differences are control MAE minus quantum-residual MAE, in $pEC_{50}$ units: positive values favor the quantum residual (right), and negative values favor the classical control (left). Dashed lines mark zero difference. Circles denote individual-seed differences, diamonds denote means, and error bars show sample standard deviations. Parentheses indicate trainable residual-head parameter counts; the quantum residual head has 36 parameters.

Relative to tuned Chemprop, quantum-residual models with both encoders achieved lower mean MAE under both ADRB2 split schemes and the MC4R structure-group split. Under the MC4R random split, tuned Chemprop had lower mean MAE than the MLP-based quantum-residual model but higher mean MAE than its GNN-based counterpart (Table 4). Complete results for all residual controls, RMSE, and $R^2$ are provided in Supporting Information Table S5, with paired comparisons in Table S6a.

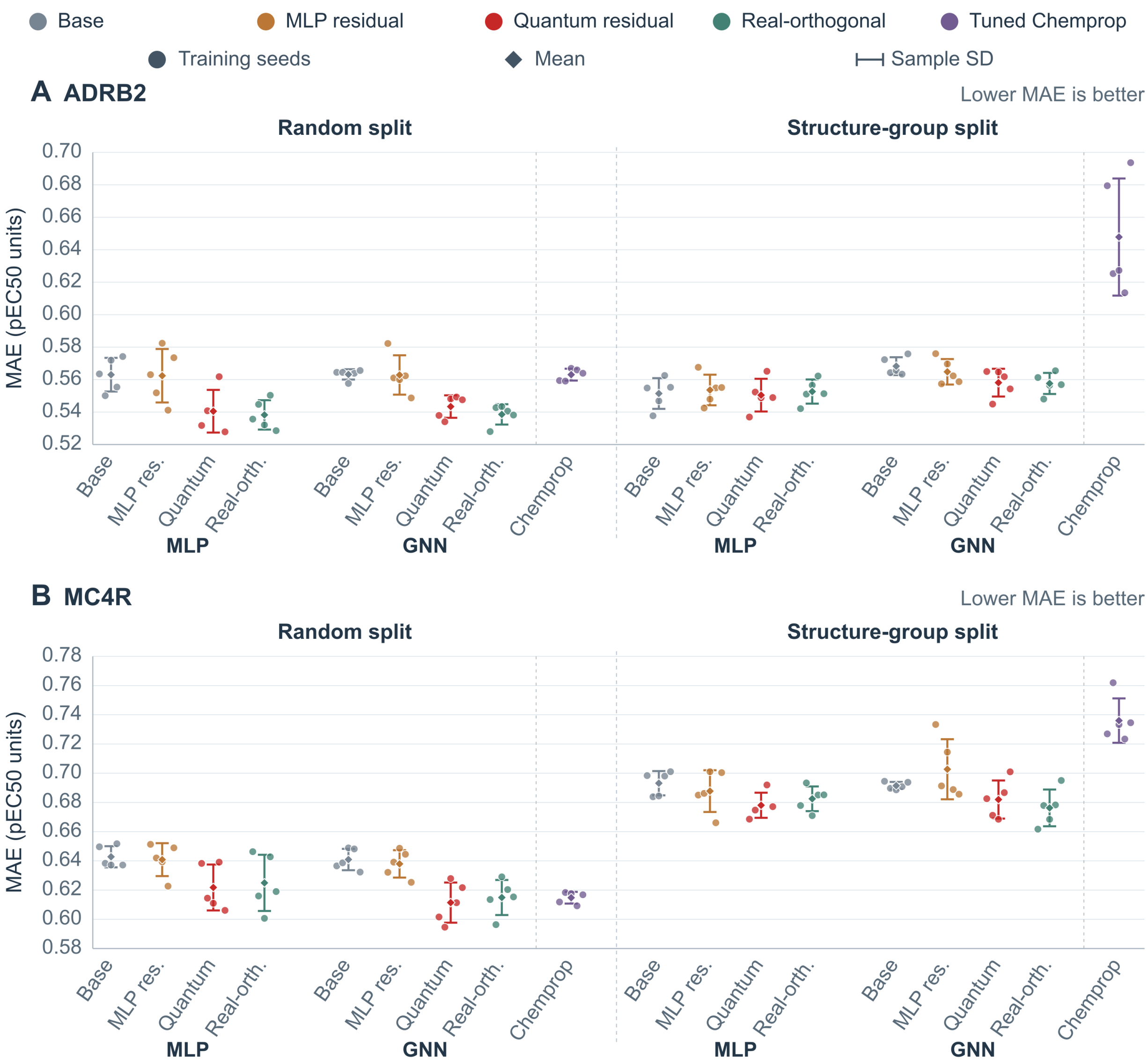

Figure 4: Functional potency prediction for ADRB2 and MC4R. (A,B) Out-of-fold MAE for ADRB2 and MC4R, respectively, under both split schemes; lower values indicate better performance. MAE axes are scaled separately for each receptor. Circles denote individual-seed results, diamonds denote means, and error bars show sample standard deviations. MLP/GNN denote encoders; MLP res. and Real-orth. denote the MLP and real-orthogonal quadratic residual heads, respectively.

# 4 Discussion

## 4.1 Predictive Gains and Structural Interpretation

The observed MAE reductions support compact quantum-circuit residual heads as a design option for receptor functional potency prediction. On GLP-1R, these gains were obtained

Table 4: Key model results and paired improvements for ADRB2 and MC4R.

| Split | Encoder/ reference | Base MAE | Quantum residual MAE | Real-orthogonal residual MAE | Tuned Chemprop MAE | Paired improvement |
|---|---|---|---|---|---|---|
| **ADRB2** | | | | | | |
| Random | MLP | $0.563 \pm 0.010$ | $0.541 \pm 0.013$ | $0.538 \pm 0.009$ | — | $0.023 \pm 0.021$ |
| | GNN | $0.563 \pm 0.003$ | $0.543 \pm 0.007$ | $0.539 \pm 0.006$ | — | $0.020 \pm 0.009$ |
| | Chemprop | — | — | — | $0.563 \pm 0.004$ | — |
| Structure-group | MLP | $0.551 \pm 0.009$ | $0.550 \pm 0.010$ | $0.553 \pm 0.007$ | — | $0.001 \pm 0.012$ |
| | GNN | $0.568 \pm 0.006$ | $0.558 \pm 0.009$ | $0.558 \pm 0.006$ | — | $0.010 \pm 0.010$ |
| | Chemprop | — | — | — | $0.648 \pm 0.036$ | — |
| **MC4R** | | | | | | |
| Random | MLP | $0.643 \pm 0.007$ | $0.622 \pm 0.016$ | $0.625 \pm 0.019$ | — | $0.021 \pm 0.009$ |
| | GNN | $0.641 \pm 0.007$ | $0.611 \pm 0.014$ | $0.615 \pm 0.012$ | — | $0.029 \pm 0.013$ |
| | Chemprop | — | — | — | $0.615 \pm 0.004$ | — |
| Structure-group | MLP | $0.693 \pm 0.008$ | $0.678 \pm 0.009$ | $0.682 \pm 0.008$ | — | $0.015 \pm 0.011$ |
| | GNN | $0.691 \pm 0.003$ | $0.682 \pm 0.013$ | $0.676 \pm 0.013$ | — | $0.009 \pm 0.011$ |
| | Chemprop | — | — | — | $0.736 \pm 0.015$ | — |

Values are means ± sample standard deviations across training seeds, in $pEC_{50}$ units. Paired improvement is base-model MAE minus quantum-residual MAE; positive values indicate improvement over the base model. — indicates not applicable.

over base models that already outperformed tuned Chemprop under the evaluation protocol used here. This finding highlights prediction-head design as an additional avenue for improving molecular prediction alongside encoder development.

The functional form of a quantum prediction head depends on its data encoding and readout.[10,25] Here, real-valued amplitude encoding and expectation-value readout yield a constrained quadratic form on the normalized latent representation. The similar predictive performance of the parameter-count-matched real-orthogonal control suggests that structured quadratic feature interactions may contribute to the observed gains. In line with the use of classical controls to interpret quantum machine learning models,[15] these findings motivate investigating how circuit-imposed constraints on feature interactions influence prediction-head performance.

This structural interpretation also depends on how the latent representation is learned. In hybrid molecular models, classical encoders supply learned representations to quantum prediction modules.[12,13] Here, joint optimization allows the representation to adapt to the residual head. The quadratic formulation describes the head's action on that representation, while the observed predictive performance reflects their joint learning. Residual-head structure and its interaction with the encoder therefore warrant consideration together in future model design.

## 4.2 Receptor Tasks and Data Context

The variation in MAE gains across receptors and split schemes indicates that the benefit of a residual head depends on the prediction task and evaluation conditions. This pattern reinforces the importance of considering molecular representations, dataset characteristics, and data partitions together.[6] Dataset size, chemical composition, label distributions, and assay conditions are potential contributors to this variation and warrant systematic investigation.

Although cAMP-related $pEC_{50}$ provided a common endpoint, the aggregated labels summarize measurements obtained under varying assay conditions.[20] Incorporating assay annotations into modeling and validating predictions under standardized experimental conditions could help clarify model applicability and strengthen the connection between predicted and experimentally measured potency.

## 4.3 Evaluation Conditions and Future Validation

Random and structure-group splits provide complementary assessments under different degrees of structural overlap between training and test sets.[6,21] Further evaluation on independent chemical series, using explicit scaffold definitions, would help establish applicability to new scaffolds. Temporal splits and prospective experimental validation could further assess performance under conditions closer to compound development.[32,33]

Classical simulations support the predictive utility of the quantum-circuit parameterization examined here. Noise-sensitivity results are provided in Supporting Information Section S7.2, Figure S1, and Table S7b. Future hardware evaluation should account for state preparation and measurement conditions.[25,34] Task-specific quantum algorithm design in biological computing[35] also motivates exploring circuit architectures tailored to molecular prediction.

# 5 Conclusions

This study evaluated compact quantum-circuit residual heads for receptor functional potency prediction, with GLP-1R cAMP-related $pEC_{50}$ as the primary task. For GLP-1R, quantum-residual models achieved lower mean MAE than the corresponding base and MLP-residual models with both encoders under both split schemes. Improvements over the base models were also observed for ADRB2 and MC4R, with magnitudes varying across tasks and evaluation conditions.

The similar predictive performance of the parameter-count-matched real-orthogonal control suggests that structured quadratic feature interactions may contribute to the gains from quantum residuals. These findings provide empirical guidance for compact residual-head design and support further exploration of quantum-circuit parameterization for molecular potency prediction.

# Data and Code Availability

Data and code are available at `https://github.com/SpinQTech/Quantum-Residual-Potency`. The repository provides prepared data, fixed splits, model configurations, and code for training and evaluating the models developed in this study. Five-fold checkpoints are supplied for seed 42, together with out-of-fold predictions for all five training seeds. Chemprop baseline predictions are included for recalculating the reported metrics. The results workbook and the scope of the accompanying materials are described in Supporting Information Section S8.

## Supporting Information

Data curation, model and training details, complete prediction metrics, paired and subgroup analyses, and simulated-noise evaluation (PDF); Supplementary Data S1: model metrics, paired comparisons, subgroup results, and noise results (XLSX).

## Author Contributions

Chenyan Jia and Cong Guo jointly led the study and designed the research. Chenyan Jia constructed the models, conducted the computational experiments and residual-structure analyses, prepared the figures, and drafted the manuscript. Cong Guo guided the experimental design, analyzed the experimental results, and supervised the research. Yaping Ma and Yuen Wang contributed to bioactivity data collection and curation, experimental planning, and the implementation of portions of the experimental code. Cong Guo, Yaping Ma, and Yuen Wang reviewed and revised the manuscript.

## Conflicts of Interest

The authors declare no competing financial interest.

## Acknowledgement

This work was supported by the Shenzhen Science and Technology Program, China (Grant No. JCYJ20241202123906009).

AI-assisted tools were used for language editing and translation.

# Supporting Information: Quantum-Circuit Residuals for Receptor Functional Potency Prediction: Predictive Gains and Quadratic Structure

Chenyan Jia,[†] Yaping Ma,[‡] Yuen Wang,[‡] and Cong Guo[*,†]

†*Shenzhen SpinQ Technology Co., Ltd., Shenzhen 518048, China*
‡*Shenzhen DIVBIO Pharmaceutical Co., Ltd., Shenzhen 518057, China*

E-mail: cguo@spinq.com

## S1 Data Sources, Curation, and Label Aggregation

ChEMBL 37 supplied cAMP-related $EC_{50}$ records for human GLP-1R (CHEMBL1784), ADRB2 (CHEMBL210), and MC4R (CHEMBL259).[S1] Sequential filtering and aggregation counts are reported in Table S1a; eligibility criteria are described in Section 2.1 of the main manuscript.

Table S1a: Sample counts after data filtering and aggregation.

| Filtering or aggregation step | GLP-1R | ADRB2 | MC4R |
|---|---|---|---|
| Original $EC_{50}$ records for the human target receptor | 2365 | 1897 | 2081 |
| Retain cAMP-related assays | 1833 | 1396 | 1208 |
| Retain the exact relation (=) | 1771 | 1083 | 1095 |
| Retain valid concentrations and convertible units | 1771 | 1077 | 1091 |
| Exclude records with data-validity issues | 1598 | 1050 | 1089 |
| Exclude records flagged as potential duplicates | 1587 | 909 | 1045 |
| Retain records with structures that can be standardized | 1577 | 909 | 1045 |
| Exclude explicit antagonism or allosteric modulation | 1560 | 909 | 928 |
| Exclude explicit receptor variants | 1147 | 909 | 886 |
| Modeling compounds after aggregation by the full InChIKey of the standardized molecular structure | 1052 | 829 | 767 |

Counts refer to activity records after successive filtering steps; the final row reports unique modeling compounds after aggregation. All eligible records had $pEC_{50}$ values between 4 and 11.

Structures were standardized with RDKit while preserving stereochemistry.[S2] Within each target, records sharing the full InChIKey of the standardized structure were assigned their median $pEC_{50}$ as the prediction label. The $pEC_{50}$ range among repeated measurements exceeded one unit for 38 GLP-1R, 14 ADRB2, and 46 MC4R compounds; these compounds were retained with median-aggregated labels.

# S2 Data Splits and Structural Connected Components

Structural connected components were defined using chirality-aware Morgan fingerprints (radius 2, 2048 bits).[S3] Molecules were linked if their Tanimoto similarity was $\geq 0.90$ or they shared the same InChIKey connectivity block. Directly or indirectly linked molecules formed a component that was kept intact during splitting. Component statistics are reported in Table S2a.

Table S2a: Summary of structural connected components.

| Target | Components | Compounds in the largest component | Singleton components |
|---|---|---|---|
| GLP-1R | 494 | 119 | 392 |
| ADRB2 | 681 | 16 | 596 |
| MC4R | 527 | 21 | 431 |

Singleton components contain one compound.

Splits were balanced by sample count and ten $pEC_{50}$ quantile bins. Random splitting assigned individual compounds, whereas structure-group splitting assigned intact connected components using mixed-integer programming. Five-fold partitions targeted training:validation:test proportions of 70:10:20 and were fixed using split seed 42 for all models within each target. Training seeds 42–46 were evaluated on these same partitions. Test-to-training nearest-neighbor similarities are summarized in Table S6c.

# S3 Models and Residual Structures

## S3.1 Classical Encoders

All encoders produce a 32-dimensional real-valued representation shared by the linear base and residual heads. The MLP maps a 256-bit Morgan fingerprint through two 64-unit hidden layers with GELU activations and a 32-unit output layer with tanh activation.

The GNN uses directed bond message passing with 33-dimensional atom features, six-dimensional bond features, a hidden dimension of 32, and a depth of 3.[S4] Message updates exclude reverse-edge contributions. Mean-pooled atom representations are concatenated with a 256-bit Morgan fingerprint and mapped through a 64-unit GELU layer and a 32-unit tanh output layer.

The Transformer processes character-level SMILES sequences of up to 1040 characters using convolutional patches of length 4 and sinusoidal positional encodings. It has two encoder layers, four attention heads, model and feedforward dimensions of 32 and 64, respectively, and dropout of 0.1. Outputs at valid positions are mean-pooled and mapped to a 32-unit tanh output layer. Prediction results are reported in Table S5 and discussed in Section S6.5.

## S3.2 Quantum Circuit and Quadratic-Form Derivation

The encoder output $h \in \mathbb{R}^{32}$ is normalized to unit Euclidean norm, giving $\widehat{h}$, and amplitude-encoded into five qubits.[S5] Each of two circuit layers applies $R_X$, $R_Y$, and $R_Z$ rotations to every qubit, followed by CNOT gates in the control-to-target sequence $0 \to 1 \to 2 \to 3 \to 4 \to 0$. The 30 rotation angles are initialized from a normal distribution with mean 0 and standard deviation 0.05. A weighted sum of the five single-qubit Pauli-$Z$ expectation values plus a bias gives the residual.

Let $U_\theta$ be the circuit unitary parameterized by rotation angles $\theta$. The readout operator is

$$D = \sum_{k=0}^{4} a_k Z_k. \tag{S1}$$

Here, $k$ indexes the qubits, $a_k$ is a trainable readout weight, and $Z_k$ acts as Pauli-$Z$ on qubit $k$ and as the identity on the other qubits, with eigenvalues $+1$ and $-1$ for states 0 and 1, respectively. For real-valued $\widehat{h}$, the antisymmetric imaginary part of the Hermitian matrix $U_\theta^\dagger D U_\theta$ contributes zero, giving

$$\Delta_\mathrm{q}(\widehat{h}) = \widehat{h}^\mathsf{T} A \widehat{h} + b. \tag{S2}$$

$$A = \mathrm{Re}(U_\theta^\dagger D U_\theta). \tag{S3}$$

Here, $\Delta_\mathrm{q}$ denotes the quantum residual, $b$ is a trainable bias, and $A$ is a $32 \times 32$ real symmetric matrix parameterized by the circuit and readout. The symbols $\dagger$, $\mathsf{T}$, and Re denote conjugate transpose, transpose, and the real part, respectively.

## S3.3 Classical Residual Controls

Each control maps the normalized representation $\widehat{h}$ to a scalar residual added to the base prediction. All weights and biases are learned independently for each residual head.

**MLP residual.** A two-unit tanh hidden layer is followed by a linear output:

$$\Delta_\mathrm{MLP}(\widehat{h}) = w_2^\mathsf{T} \tanh(W_1 \widehat{h} + b_1) + b_2. \tag{S4}$$

Here, $W_1$ and $w_2$ are the input and output weights, with dimensions $2 \times 32$ and 2, respectively; $b_1$ and $b_2$ are the corresponding vector and scalar biases. The tanh activation is applied elementwise. This head has 69 trainable parameters.

**Diagonal quadratic form.** The residual is a weighted sum of squared input features plus a bias:

$$\Delta_\mathrm{diag}(\widehat{h}) = \sum_{j=0}^{31} c_j \widehat{h}_j^2 + b. \tag{S5}$$

Here, $\widehat{h}_j$ is input feature $j$, $c_j$ is its squared-term weight, and $b$ is a scalar bias. This head has 33 trainable parameters.

**Low-rank quadratic form.** The input is linearly projected onto two features, whose squares and cross product are combined in a weighted sum:

$$\Delta_{\mathrm{rank2}}(\widehat{h}) = \widehat{h}^{\mathsf{T}} W S W^{\mathsf{T}} \widehat{h} + b. \tag{S6}$$

Here, $W \in \mathbb{R}^{32\times 2}$ projects the input onto two features through $W^{\mathsf{T}}\widehat{h}$. The symmetric $2 \times 2$ matrix $S$ contains three trainable coefficients for their squared terms and cross term. The resulting coefficient matrix $WSW^{\mathsf{T}}$ has rank at most 2. Including the scalar bias $b$, this head has 68 trainable parameters.

**Real-orthogonal quadratic form.** The real-orthogonal control comprises two layers of trainable rotations and fixed coordinate permutations that preserve the input norm. Each layer uses 15 independent angles, each shared across a predefined set of coordinate pairs. Each pair is transformed as follows:

$$\begin{pmatrix} v_i' \\ v_j' \end{pmatrix} = \begin{pmatrix} \cos\alpha & -\sin\alpha \\ \sin\alpha & \cos\alpha \end{pmatrix} \begin{pmatrix} v_i \\ v_j \end{pmatrix}. \tag{S7}$$

Here, $v_i$ and $v_j$ denote the coordinates before rotation, primes denote their transformed values, and $\alpha$ is a trainable angle; cos and sin are the cosine and sine functions. Each layer applies rotations followed by fixed coordinate permutations matching the quantum circuit's CNOT sequence. Together, the two layers define a $32 \times 32$ real orthogonal matrix $Q$. The residual is a weighted sum of the squared coordinates of $Q\widehat{h}$, with weights specified by the diagonal readout matrix $D$ defined in Section S3.2:

$$\Delta_{\mathrm{orth}}(\widehat{h}) = \widehat{h}^{\mathsf{T}} Q^{\mathsf{T}} D Q \widehat{h} + b. \tag{S8}$$

Here, orth denotes the real-orthogonal control, and $b$ is a scalar bias. The readout weights in $D$ are learned independently for this model. Its 30 rotation angles, five readout weights, and one bias give 36 trainable parameters, matching the quantum residual. The two heads also share the same layer count, fixed coordinate permutations, and readout structure.

Each residual model jointly optimizes its encoder, linear base head, and residual head. Performance comparisons therefore reflect both the residual structure and the representation learned during joint training.

# S4 Training and Chemprop Tuning

## S4.1 Training and Model Selection

Within each fold, $\mathrm{pEC}_{50}$ labels were standardized using the training-set mean and population standard deviation for training, validation, and testing. Predictions were transformed back to the original $\mathrm{pEC}_{50}$ scale before evaluation.

Models were implemented in PyTorch and trained using AdamW with cosine learning-rate decay. The batch-averaged Smooth L1 loss used a transition threshold of 0.5, and gradient norms were clipped to 5. The encoder and base head shared a learning rate, while the residual head used a separate rate. Encoder-specific settings are listed below.

**Training configurations.**

| Setting | MLP | GNN | Transformer |
|---|---|---|---|
| Maximum epochs | 100 | 35 | 60 |
| Batch size | 64 | 32 | 32 |
| Encoder/base-head learning rate | 0.001 | 0.0005 | 0.0005 |
| Residual-head learning rate | 0.003 | 0.001 | 0.001 |
| Weight decay | 0.0001 | 0.00001 | 0.0001 |
| Early-stopping patience (epochs) | 12 | 8 | 10 |
| Minimum epochs | 15 | 10 | 12 |

All variants using the same encoder shared these settings. After the minimum training duration, early stopping was triggered when validation MAE failed to improve for the specified patience period, subject to the maximum epoch limit. The checkpoint with the lowest validation MAE in each fold was used for test-set prediction.

### S4.2 Chemprop Training and Hyperparameter Selection

Chemprop 2.3.1 used the same compounds and data splits as the other models.[S6] A grid search evaluated 12 configurations combining message-passing depths of {2, 3, 4}, hidden dimensions of {300, 600}, and dropout rates of {0, 0.2}. The hidden dimension was shared by the message-passing and feedforward layers. Training used a maximum of 100 epochs, two warmup epochs, and early stopping with a patience of 12 epochs based on validation MAE.

For each target and split scheme, the configuration with the lowest mean validation MAE across folds at seed 42 was selected and fixed for evaluation with seeds 42–46 (Table S4a). Configuration selection and out-of-fold evaluation therefore share the same cross-validation partitions. Chemprop comparisons reflect this selection protocol, whereas residual gains are assessed through paired comparisons within each encoder architecture.

Two additional Chemprop settings were evaluated. The default setting used a message-passing depth of 3, a hidden dimension of 300, zero dropout, and 50 training epochs. The training-protocol-adjusted setting retained this architecture but adopted the epoch limit, warmup, and early-stopping settings used in the grid search. Table S5b compares MAE across the default, training-protocol-adjusted, and tuned settings.

## S5 Prediction Metrics

Table S5 reports MAE, RMSE, and $R^2$ for the base models, residual models, and tuned Chemprop by target and split scheme. Metrics were calculated from pooled out-of-fold predictions for each training seed and summarized as the mean ± sample standard deviation across seeds. The standard deviation reflects training variability under fixed data splits.

Table S4a: Selected Chemprop configurations.

| Target | Split | Message-passing depth | Message-passing/feedforward hidden dimension | Dropout |
|---|---|---|---|---|
| GLP-1R | Random | 3 | 600 | 0.2 |
| GLP-1R | Structure-group | 3 | 600 | 0.2 |
| ADRB2 | Random | 4 | 600 | 0.2 |
| ADRB2 | Structure-group | 4 | 600 | 0 |
| MC4R | Random | 4 | 600 | 0 |
| MC4R | Structure-group | 4 | 600 | 0 |

Lower MAE and RMSE (in $pEC_{50}$ units) and higher $R^2$ indicate better predictive performance. Unrounded summary and per-seed metrics are available in the "Model Summary" and "Model by Seed" worksheets of Supplementary Data S1, respectively.

Table S5: Prediction metrics for the three receptor tasks (mean ± sample standard deviation).

| Encoder/baseline | Model variant | MAE | RMSE | $R^2$ |
|---|---|---|---|---|
| **GLP-1R — Random split** | | | | |
| MLP | Base | 0.528 ± 0.006 | 0.727 ± 0.008 | 0.759 ± 0.006 |
| MLP | MLP residual | 0.521 ± 0.007 | 0.728 ± 0.008 | 0.758 ± 0.005 |
| MLP | Quantum residual | 0.504 ± 0.008 | 0.713 ± 0.012 | 0.768 ± 0.008 |
| MLP | Diagonal quadratic | 0.528 ± 0.003 | 0.729 ± 0.007 | 0.757 ± 0.004 |
| MLP | Low-rank quadratic | 0.524 ± 0.005 | 0.727 ± 0.005 | 0.759 ± 0.003 |
| MLP | Real-orthogonal quadratic | 0.509 ± 0.015 | 0.722 ± 0.018 | 0.762 ± 0.012 |
| GNN | Base | 0.550 ± 0.008 | 0.752 ± 0.010 | 0.742 ± 0.007 |
| GNN | MLP residual | 0.536 ± 0.007 | 0.739 ± 0.007 | 0.751 ± 0.005 |
| GNN | Quantum residual | 0.511 ± 0.011 | 0.719 ± 0.012 | 0.764 ± 0.008 |
| GNN | Diagonal quadratic | 0.545 ± 0.006 | 0.746 ± 0.007 | 0.746 ± 0.005 |
| GNN | Low-rank quadratic | 0.549 ± 0.008 | 0.752 ± 0.008 | 0.742 ± 0.006 |
| GNN | Real-orthogonal quadratic | 0.508 ± 0.003 | 0.717 ± 0.007 | 0.766 ± 0.004 |
| Transformer | Base | 0.677 ± 0.006 | 0.939 ± 0.012 | 0.598 ± 0.010 |
| Transformer | MLP residual | 0.683 ± 0.004 | 0.947 ± 0.013 | 0.590 ± 0.011 |
| Transformer | Quantum residual | 0.680 ± 0.005 | 0.939 ± 0.007 | 0.598 ± 0.006 |
| Chemprop | 2.3.1 tuned | 0.577 ± 0.015 | 0.769 ± 0.015 | 0.730 ± 0.010 |
| **GLP-1R — Structure-group split** | | | | |
| MLP | Base | 0.677 ± 0.010 | 0.931 ± 0.015 | 0.604 ± 0.012 |
| MLP | MLP residual | 0.660 ± 0.017 | 0.905 ± 0.020 | 0.626 ± 0.016 |
| MLP | Quantum residual | 0.639 ± 0.016 | 0.887 ± 0.028 | 0.641 ± 0.023 |
| MLP | Diagonal quadratic | 0.664 ± 0.014 | 0.921 ± 0.013 | 0.613 ± 0.011 |
| MLP | Low-rank quadratic | 0.670 ± 0.014 | 0.926 ± 0.017 | 0.609 ± 0.014 |
| MLP | Real-orthogonal quadratic | 0.659 ± 0.022 | 0.910 ± 0.029 | 0.621 ± 0.025 |
| GNN | Base | 0.680 ± 0.009 | 0.925 ± 0.010 | 0.609 ± 0.009 |
| GNN | MLP residual | 0.669 ± 0.007 | 0.913 ± 0.012 | 0.619 ± 0.010 |
| GNN | Quantum residual | 0.641 ± 0.027 | 0.886 ± 0.032 | 0.642 ± 0.026 |
| GNN | Diagonal quadratic | 0.674 ± 0.008 | 0.923 ± 0.009 | 0.611 ± 0.007 |
| GNN | Low-rank quadratic | 0.678 ± 0.009 | 0.925 ± 0.012 | 0.609 ± 0.010 |
| GNN | Real-orthogonal quadratic | 0.649 ± 0.029 | 0.897 ± 0.034 | 0.632 ± 0.028 |
| Transformer | Base | 0.718 ± 0.014 | 0.961 ± 0.018 | 0.578 ± 0.016 |
| Transformer | MLP residual | 0.716 ± 0.016 | 0.959 ± 0.020 | 0.580 ± 0.018 |
| Transformer | Quantum residual | 0.708 ± 0.023 | 0.954 ± 0.033 | 0.584 ± 0.029 |
| Chemprop | 2.3.1 tuned | 0.737 ± 0.050 | 0.982 ± 0.056 | 0.559 ± 0.051 |

Continued on next page

Table S5: Prediction metrics for the three receptor tasks (continued).

| Encoder/baseline | Model variant | MAE | RMSE | $R^2$ |
|---|---|---|---|---|
| **ADRB2 — Random split** | | | | |
| MLP | Base | 0.563 ± 0.010 | 0.794 ± 0.010 | 0.683 ± 0.008 |
| MLP | MLP residual | 0.562 ± 0.016 | 0.797 ± 0.016 | 0.681 ± 0.012 |
| MLP | Quantum residual | 0.541 ± 0.013 | 0.779 ± 0.009 | 0.695 ± 0.007 |
| MLP | Diagonal quadratic | 0.563 ± 0.017 | 0.797 ± 0.016 | 0.680 ± 0.013 |
| MLP | Low-rank quadratic | 0.563 ± 0.010 | 0.795 ± 0.009 | 0.682 ± 0.008 |
| MLP | Real-orthogonal quadratic | 0.538 ± 0.009 | 0.774 ± 0.008 | 0.699 ± 0.006 |
| GNN | Base | 0.563 ± 0.003 | 0.791 ± 0.002 | 0.686 ± 0.001 |
| GNN | MLP residual | 0.563 ± 0.012 | 0.788 ± 0.013 | 0.688 ± 0.010 |
| GNN | Quantum residual | 0.543 ± 0.007 | 0.774 ± 0.014 | 0.699 ± 0.011 |
| GNN | Diagonal quadratic | 0.560 ± 0.006 | 0.789 ± 0.004 | 0.687 ± 0.003 |
| GNN | Low-rank quadratic | 0.563 ± 0.003 | 0.791 ± 0.001 | 0.686 ± 0.001 |
| GNN | Real-orthogonal quadratic | 0.539 ± 0.006 | 0.764 ± 0.008 | 0.706 ± 0.006 |
| Transformer | Base | 0.801 ± 0.028 | 1.119 ± 0.029 | 0.370 ± 0.033 |
| Transformer | MLP residual | 0.806 ± 0.020 | 1.125 ± 0.025 | 0.363 ± 0.028 |
| Transformer | Quantum residual | 0.800 ± 0.036 | 1.114 ± 0.040 | 0.376 ± 0.046 |
| Chemprop | 2.3.1 tuned | 0.563 ± 0.004 | 0.798 ± 0.006 | 0.680 ± 0.005 |
| **ADRB2 — Structure-group split** | | | | |
| MLP | Base | 0.551 ± 0.009 | 0.797 ± 0.009 | 0.681 ± 0.008 |
| MLP | MLP residual | 0.554 ± 0.009 | 0.800 ± 0.011 | 0.678 ± 0.009 |
| MLP | Quantum residual | 0.550 ± 0.010 | 0.805 ± 0.005 | 0.675 ± 0.004 |
| MLP | Diagonal quadratic | 0.552 ± 0.001 | 0.797 ± 0.004 | 0.681 ± 0.003 |
| MLP | Low-rank quadratic | 0.554 ± 0.007 | 0.798 ± 0.008 | 0.680 ± 0.006 |
| MLP | Real-orthogonal quadratic | 0.553 ± 0.007 | 0.807 ± 0.009 | 0.673 ± 0.007 |
| GNN | Base | 0.568 ± 0.006 | 0.800 ± 0.005 | 0.678 ± 0.004 |
| GNN | MLP residual | 0.565 ± 0.008 | 0.797 ± 0.008 | 0.681 ± 0.006 |
| GNN | Quantum residual | 0.558 ± 0.009 | 0.804 ± 0.014 | 0.675 ± 0.011 |
| GNN | Diagonal quadratic | 0.564 ± 0.006 | 0.796 ± 0.005 | 0.682 ± 0.004 |
| GNN | Low-rank quadratic | 0.568 ± 0.005 | 0.801 ± 0.005 | 0.678 ± 0.004 |
| GNN | Real-orthogonal quadratic | 0.558 ± 0.006 | 0.802 ± 0.014 | 0.677 ± 0.011 |
| Transformer | Base | 0.833 ± 0.020 | 1.127 ± 0.021 | 0.362 ± 0.023 |
| Transformer | MLP residual | 0.839 ± 0.015 | 1.136 ± 0.014 | 0.351 ± 0.016 |
| Transformer | Quantum residual | 0.831 ± 0.036 | 1.132 ± 0.036 | 0.356 ± 0.041 |
| Chemprop | 2.3.1 tuned | 0.648 ± 0.036 | 0.889 ± 0.043 | 0.602 ± 0.038 |

Continued on next page

Table S5: Prediction metrics for the three receptor tasks (continued).

| Encoder/baseline | Model variant | MAE | RMSE | $R^2$ |
|---|---|---|---|---|
| **MC4R — Random split** | | | | |
| MLP | Base | 0.643 ± 0.007 | 0.822 ± 0.010 | 0.462 ± 0.012 |
| MLP | MLP residual | 0.641 ± 0.011 | 0.820 ± 0.016 | 0.464 ± 0.020 |
| MLP | Quantum residual | 0.622 ± 0.016 | 0.803 ± 0.017 | 0.487 ± 0.021 |
| MLP | Diagonal quadratic | 0.645 ± 0.007 | 0.825 ± 0.012 | 0.459 ± 0.015 |
| MLP | Low-rank quadratic | 0.644 ± 0.006 | 0.826 ± 0.011 | 0.457 ± 0.015 |
| MLP | Real-orthogonal quadratic | 0.625 ± 0.019 | 0.808 ± 0.021 | 0.480 ± 0.027 |
| GNN | Base | 0.641 ± 0.007 | 0.819 ± 0.009 | 0.466 ± 0.011 |
| GNN | MLP residual | 0.638 ± 0.009 | 0.817 ± 0.010 | 0.469 ± 0.013 |
| GNN | Quantum residual | 0.611 ± 0.014 | 0.785 ± 0.016 | 0.509 ± 0.020 |
| GNN | Diagonal quadratic | 0.635 ± 0.004 | 0.810 ± 0.006 | 0.478 ± 0.007 |
| GNN | Low-rank quadratic | 0.640 ± 0.008 | 0.818 ± 0.010 | 0.467 ± 0.012 |
| GNN | Real-orthogonal quadratic | 0.615 ± 0.012 | 0.792 ± 0.010 | 0.500 ± 0.013 |
| Transformer | Base | 0.760 ± 0.018 | 0.978 ± 0.019 | 0.239 ± 0.029 |
| Transformer | MLP residual | 0.754 ± 0.006 | 0.971 ± 0.008 | 0.250 ± 0.013 |
| Transformer | Quantum residual | 0.717 ± 0.016 | 0.930 ± 0.015 | 0.311 ± 0.023 |
| Chemprop | 2.3.1 tuned | 0.615 ± 0.004 | 0.803 ± 0.004 | 0.487 ± 0.006 |
| **MC4R — Structure-group split** | | | | |
| MLP | Base | 0.693 ± 0.008 | 0.887 ± 0.007 | 0.373 ± 0.010 |
| MLP | MLP residual | 0.688 ± 0.014 | 0.883 ± 0.019 | 0.379 ± 0.026 |
| MLP | Quantum residual | 0.678 ± 0.009 | 0.876 ± 0.016 | 0.389 ± 0.022 |
| MLP | Diagonal quadratic | 0.690 ± 0.004 | 0.884 ± 0.008 | 0.377 ± 0.011 |
| MLP | Low-rank quadratic | 0.697 ± 0.006 | 0.891 ± 0.007 | 0.368 ± 0.010 |
| MLP | Real-orthogonal quadratic | 0.682 ± 0.008 | 0.881 ± 0.014 | 0.382 ± 0.020 |
| GNN | Base | 0.691 ± 0.003 | 0.882 ± 0.005 | 0.381 ± 0.007 |
| GNN | MLP residual | 0.703 ± 0.021 | 0.895 ± 0.021 | 0.362 ± 0.031 |
| GNN | Quantum residual | 0.682 ± 0.013 | 0.879 ± 0.015 | 0.386 ± 0.021 |
| GNN | Diagonal quadratic | 0.687 ± 0.004 | 0.878 ± 0.006 | 0.387 ± 0.009 |
| GNN | Low-rank quadratic | 0.691 ± 0.002 | 0.881 ± 0.005 | 0.382 ± 0.008 |
| GNN | Real-orthogonal quadratic | 0.676 ± 0.013 | 0.872 ± 0.017 | 0.395 ± 0.023 |
| Transformer | Base | 0.873 ± 0.024 | 1.090 ± 0.013 | 0.054 ± 0.023 |
| Transformer | MLP residual | 0.873 ± 0.021 | 1.090 ± 0.020 | 0.055 ± 0.035 |
| Transformer | Quantum residual | 0.846 ± 0.023 | 1.061 ± 0.022 | 0.103 ± 0.037 |
| Chemprop | 2.3.1 tuned | 0.736 ± 0.015 | 0.939 ± 0.018 | 0.299 ± 0.026 |

Table S5b: MAE for Chemprop 2.3.1 under different training and tuning settings (mean ± sample standard deviation).

| Target | Split | Default reference | Training-protocol-adjusted reference | Tuned model |
|---|---|---|---|---|
| GLP-1R | Random | 0.673 ± 0.004 | 0.610 ± 0.007 | 0.577 ± 0.015 |
| GLP-1R | Structure-group | 0.719 ± 0.006 | 0.768 ± 0.050 | 0.737 ± 0.050 |
| ADRB2 | Random | 0.751 ± 0.018 | 0.622 ± 0.005 | 0.563 ± 0.004 |
| ADRB2 | Structure-group | 0.760 ± 0.021 | 0.735 ± 0.007 | 0.648 ± 0.036 |
| MC4R | Random | 0.737 ± 0.006 | 0.676 ± 0.018 | 0.615 ± 0.004 |
| MC4R | Structure-group | 0.832 ± 0.008 | 0.809 ± 0.026 | 0.736 ± 0.015 |

The training-protocol-adjusted setting changes only the training schedule, whereas the tuned setting also uses validation-selected hyperparameters (Section S4.2).

# S6 Paired Analyses and Subgroup Results

## S6.1 Metrics and Aggregation across Training Seeds

For each target, let $N$ be the number of compounds, $y_i$ the observed $\mathrm{pEC}_{50}$ of compound $i$, $\widehat{y}_{i,s}$ its out-of-fold prediction for training seed $s$, and $\overline{y}$ the mean observed $\mathrm{pEC}_{50}$. Each compound contributes once to the metrics:

$$\mathrm{MAE}_s = \frac{1}{N}\sum_{i=1}^{N} |\widehat{y}_{i,s} - y_i|. \tag{S9}$$

$$\mathrm{RMSE}_s = \sqrt{\frac{1}{N}\sum_{i=1}^{N} (\widehat{y}_{i,s} - y_i)^2}. \tag{S10}$$

$$R_s^2 = 1 - \frac{\sum_{i=1}^{N}(\widehat{y}_{i,s} - y_i)^2}{\sum_{i=1}^{N}(y_i - \overline{y})^2}. \tag{S11}$$

Because training sets overlap across folds,[S7] metrics were summarized across training seeds using each seed's complete out-of-fold predictions. MAE differences were paired by target, split scheme, encoder, and training seed, then summarized as the mean ± sample standard deviation across seeds. Differences were defined as control MAE minus quantum-residual MAE; positive values indicate an improvement with the quantum residual.

## S6.2 Paired Bootstrap over Structural Connected Components

Paired bootstrap resampling used the structural connected components defined in Section S2 as sampling units.[S8] Within each target, split scheme, encoder, and training seed, components were sampled uniformly with replacement, retaining all constituent compounds. Each replicate contained the original number of components, and both models were evaluated on the same resampled compounds using their fixed out-of-fold predictions. MAE differences were calculated across compounds, preserving within-component grouping.

For each training seed, 95% bootstrap intervals were calculated from the 2.5th and 97.5th percentiles of 10,000 resampled MAE differences. These intervals describe evaluation-sample variability conditional on fixed predictions and structural grouping. They are reported descriptively, without multiplicity adjustment, in the “Paired Comparisons” worksheet of Supplementary Data S1.

## S6.3 Key Paired Comparisons

Table S6a reports paired MAE differences between quantum-residual models and the corresponding base models or parameter-count-matched real-orthogonal controls within each encoder architecture. Positive values indicate an improvement with the quantum residual. Individual model metrics are reported in Table S5.

## S6.4 Amide-Count and Training-Neighbor Subgroups

Subgroup analyses used out-of-fold predictions from models trained on the full training sets for each target. Table S6b groups compounds by amide-bond count ($\leq 2$ or $\geq 3$), an operational distinction describing structural composition rather than a formal chemical classification.

Table S6c groups test compounds by their maximum Tanimoto similarity to the corresponding training set ($< 0.90$ or $\geq 0.90$), using the fingerprints described in Section S2. Under structure-group splitting, all test compounds fell below 0.90.

Both tables compare base and quantum-residual models with MLP and GNN encoders; $n$ denotes the number of compounds in each subgroup. Corresponding metrics are available in the “Subgroup Results” worksheet of Supplementary Data S1.

Table S6a: Paired MAE differences: control minus quantum-residual model (mean ± sample standard deviation).

| Target | Split | Encoder | Control | MAE difference |
|---|---|---|---|---|
| GLP-1R | Random | MLP | Base | 0.024 ± 0.011 |
| GLP-1R | Random | MLP | Real-orthogonal quadratic | 0.005 ± 0.012 |
| GLP-1R | Random | GNN | Base | 0.039 ± 0.013 |
| GLP-1R | Random | GNN | Real-orthogonal quadratic | -0.003 ± 0.011 |
| GLP-1R | Structure-group | MLP | Base | 0.038 ± 0.009 |
| GLP-1R | Structure-group | MLP | Real-orthogonal quadratic | 0.020 ± 0.011 |
| GLP-1R | Structure-group | GNN | Base | 0.039 ± 0.031 |
| GLP-1R | Structure-group | GNN | Real-orthogonal quadratic | 0.008 ± 0.008 |
| ADRB2 | Random | MLP | Base | 0.023 ± 0.021 |
| ADRB2 | Random | MLP | Real-orthogonal quadratic | -0.002 ± 0.007 |
| ADRB2 | Random | GNN | Base | 0.020 ± 0.009 |
| ADRB2 | Random | GNN | Real-orthogonal quadratic | -0.005 ± 0.008 |
| ADRB2 | Structure-group | MLP | Base | 0.001 ± 0.012 |
| ADRB2 | Structure-group | MLP | Real-orthogonal quadratic | 0.002 ± 0.004 |
| ADRB2 | Structure-group | GNN | Base | 0.010 ± 0.010 |
| ADRB2 | Structure-group | GNN | Real-orthogonal quadratic | 0.000 ± 0.005 |
| MC4R | Random | MLP | Base | 0.021 ± 0.009 |
| MC4R | Random | MLP | Real-orthogonal quadratic | 0.003 ± 0.009 |
| MC4R | Random | GNN | Base | 0.029 ± 0.013 |
| MC4R | Random | GNN | Real-orthogonal quadratic | 0.003 ± 0.007 |
| MC4R | Structure-group | MLP | Base | 0.015 ± 0.011 |
| MC4R | Structure-group | MLP | Real-orthogonal quadratic | 0.004 ± 0.011 |
| MC4R | Structure-group | GNN | Base | 0.009 ± 0.011 |
| MC4R | Structure-group | GNN | Real-orthogonal quadratic | -0.006 ± 0.010 |

Table S6b: MAE by amide-bond count (mean ± sample standard deviation).

| Split | Amide bonds | $n$ | Encoder | Base MAE | Quantum residual MAE |
|---|---|---|---|---|---|
| **GLP-1R** | | | | | |
| Random | $\leq 2$ | 472 | MLP | 0.513 ± 0.005 | 0.490 ± 0.012 |
| Random | $\leq 2$ | 472 | GNN | 0.515 ± 0.010 | 0.497 ± 0.008 |
| Random | $\geq 3$ | 580 | MLP | 0.539 ± 0.007 | 0.514 ± 0.007 |
| Random | $\geq 3$ | 580 | GNN | 0.579 ± 0.009 | 0.523 ± 0.020 |
| Structure-group | $\leq 2$ | 472 | MLP | 0.558 ± 0.023 | 0.494 ± 0.022 |
| Structure-group | $\leq 2$ | 472 | GNN | 0.573 ± 0.014 | 0.526 ± 0.032 |
| Structure-group | $\geq 3$ | 580 | MLP | 0.773 ± 0.015 | 0.757 ± 0.019 |
| Structure-group | $\geq 3$ | 580 | GNN | 0.766 ± 0.011 | 0.734 ± 0.025 |
| **ADRB2** | | | | | |
| Random | $\leq 2$ | 711 | MLP | 0.609 ± 0.012 | 0.594 ± 0.014 |
| Random | $\leq 2$ | 711 | GNN | 0.608 ± 0.003 | 0.596 ± 0.009 |
| Random | $\geq 3$ | 118 | MLP | 0.284 ± 0.014 | 0.216 ± 0.015 |
| Random | $\geq 3$ | 118 | GNN | 0.295 ± 0.011 | 0.228 ± 0.018 |
| Structure-group | $\leq 2$ | 711 | MLP | 0.598 ± 0.011 | 0.604 ± 0.008 |
| Structure-group | $\leq 2$ | 711 | GNN | 0.607 ± 0.006 | 0.611 ± 0.010 |
| Structure-group | $\geq 3$ | 118 | MLP | 0.271 ± 0.011 | 0.225 ± 0.030 |
| Structure-group | $\geq 3$ | 118 | GNN | 0.334 ± 0.010 | 0.239 ± 0.017 |
| **MC4R** | | | | | |
| Random | $\leq 2$ | 379 | MLP | 0.582 ± 0.010 | 0.576 ± 0.015 |
| Random | $\leq 2$ | 379 | GNN | 0.576 ± 0.004 | 0.567 ± 0.018 |
| Random | $\geq 3$ | 388 | MLP | 0.702 ± 0.018 | 0.666 ± 0.019 |
| Random | $\geq 3$ | 388 | GNN | 0.705 ± 0.013 | 0.655 ± 0.012 |
| Structure-group | $\leq 2$ | 379 | MLP | 0.599 ± 0.011 | 0.582 ± 0.017 |
| Structure-group | $\leq 2$ | 379 | GNN | 0.587 ± 0.003 | 0.578 ± 0.010 |
| Structure-group | $\geq 3$ | 388 | MLP | 0.785 ± 0.009 | 0.772 ± 0.024 |
| Structure-group | $\geq 3$ | 388 | GNN | 0.794 ± 0.006 | 0.783 ± 0.021 |

Table S6c: MAE by training-set nearest-neighbor similarity (mean ± sample standard deviation).

| Split | Maximum similarity | $n$ | Encoder | Base MAE | Quantum residual MAE |
|---|---|---|---|---|---|
| **GLP-1R** | | | | | |
| Random | $\geq 0.90$ | 609 | MLP | 0.474 ± 0.005 | 0.451 ± 0.009 |
| Random | $\geq 0.90$ | 609 | GNN | 0.515 ± 0.006 | 0.453 ± 0.018 |
| Random | $< 0.90$ | 443 | MLP | 0.602 ± 0.007 | 0.576 ± 0.015 |
| Random | $< 0.90$ | 443 | GNN | 0.599 ± 0.012 | 0.591 ± 0.008 |
| Structure-group | $< 0.90$ | 1052 | MLP | 0.677 ± 0.010 | 0.639 ± 0.016 |
| Structure-group | $< 0.90$ | 1052 | GNN | 0.680 ± 0.009 | 0.641 ± 0.027 |
| **ADRB2** | | | | | |
| Random | $\geq 0.90$ | 140 | MLP | 0.355 ± 0.015 | 0.311 ± 0.015 |
| Random | $\geq 0.90$ | 140 | GNN | 0.369 ± 0.009 | 0.321 ± 0.009 |
| Random | $< 0.90$ | 689 | MLP | 0.605 ± 0.011 | 0.587 ± 0.013 |
| Random | $< 0.90$ | 689 | GNN | 0.603 ± 0.003 | 0.589 ± 0.008 |
| Structure-group | $< 0.90$ | 829 | MLP | 0.551 ± 0.009 | 0.550 ± 0.010 |
| Structure-group | $< 0.90$ | 829 | GNN | 0.568 ± 0.006 | 0.558 ± 0.009 |
| **MC4R** | | | | | |
| Random | $\geq 0.90$ | 244 | MLP | 0.631 ± 0.021 | 0.582 ± 0.022 |
| Random | $\geq 0.90$ | 244 | GNN | 0.636 ± 0.015 | 0.577 ± 0.016 |
| Random | $< 0.90$ | 523 | MLP | 0.648 ± 0.007 | 0.640 ± 0.017 |
| Random | $< 0.90$ | 523 | GNN | 0.643 ± 0.005 | 0.628 ± 0.018 |
| Structure-group | $< 0.90$ | 767 | MLP | 0.693 ± 0.008 | 0.678 ± 0.009 |
| Structure-group | $< 0.90$ | 767 | GNN | 0.691 ± 0.003 | 0.682 ± 0.013 |

Quantum-residual models achieved lower mean MAE than the corresponding base models in both amide-count groups across both encoders and split schemes for GLP-1R and MC4R (Table S6b). For ADRB2, improvements occurred in both groups under random splitting; under structure-group splitting, mean MAE decreased for compounds with $\geq 3$ amide bonds but increased slightly for those with $\leq 2$.

## S6.5 Supplementary Transformer Evaluation

Across all three targets and both split schemes, quantum-residual models with MLP and GNN encoders achieved lower mean MAE than their Transformer counterparts (Table S5). These results informed the choice of MLP and GNN for the structural-control experiments. The Transformer quantum-residual model achieved lower mean MAE than tuned Chemprop under GLP-1R structure-group splitting but higher mean MAE in the other five settings.

# S7 Inference under Simulated Noise

## S7.1 Evaluation Setup and Noise Parameters

Noise sensitivity was evaluated for GLP-1R quantum-residual models with MLP and GNN encoders under both split schemes. Circuit and readout noise were introduced only during inference, retaining the model parameters and label standardization from training under ideal conditions. Density-matrix simulations assumed ideal preparation of the amplitude-encoded input and local Markovian noise on the qubits involved in each gate. Single-qubit rotations were followed by depolarization and then amplitude damping; CNOT gates were followed by amplitude damping on both qubits and then two-qubit depolarization.

The single- and two-qubit depolarizing strengths, $p_1$ and $p_2$ (Table S7a), correspond to total nonidentity Pauli-error probabilities of $3p_1/4$ and $15p_2/16$, respectively. The amplitude-damping probability is

$$\gamma = 1 - \exp(-t/T_1). \tag{S12}$$

Here, $\gamma$ denotes the damping probability, $t$ the gate duration, $T_1$ the energy-relaxation time, and exp the exponential function. Gate durations were 40 ns for single-qubit rotations and 300 ns for CNOT gates.

The asymmetric readout model assigned misclassification probabilities of $2\varepsilon/3$ for $0 \to 1$ and $4\varepsilon/3$ for $1 \to 0$, where $\varepsilon$ is their arithmetic mean. These errors were incorporated into each qubit's Pauli-$Z$ expectation value before applying the fixed linear readout. N1–N5 represent progressively stronger simulated noise levels with device-independent parameter settings.

Table S7a: Parameters for the five simulated noise levels.

| Level | $p_1$ | $p_2$ | $T_1$ ($\mu$s) | $\varepsilon$ |
|---|---|---|---|---|
| N1 | 0.00005 | 0.0003 | 300 | 0.001 |
| N2 | 0.0001 | 0.001 | 150 | 0.003 |
| N3 | 0.0002 | 0.003 | 100 | 0.007 |
| N4 | 0.0005 | 0.01 | 50 | 0.015 |
| N5 | 0.001 | 0.03 | 20 | 0.03 |

## S7.2 Noise-Sensitivity Results

Figure S1 shows MAE across noise levels relative to the classical MLP-residual reference. Table S7b reports MAE and two paired differences: noise-induced MAE change (noisy minus ideal MAE) and improvement over the base model (base minus noisy MAE). Positive values indicate increased error and retained improvement, respectively. Differences were calculated within each training seed and summarized as the mean ± sample standard deviation.

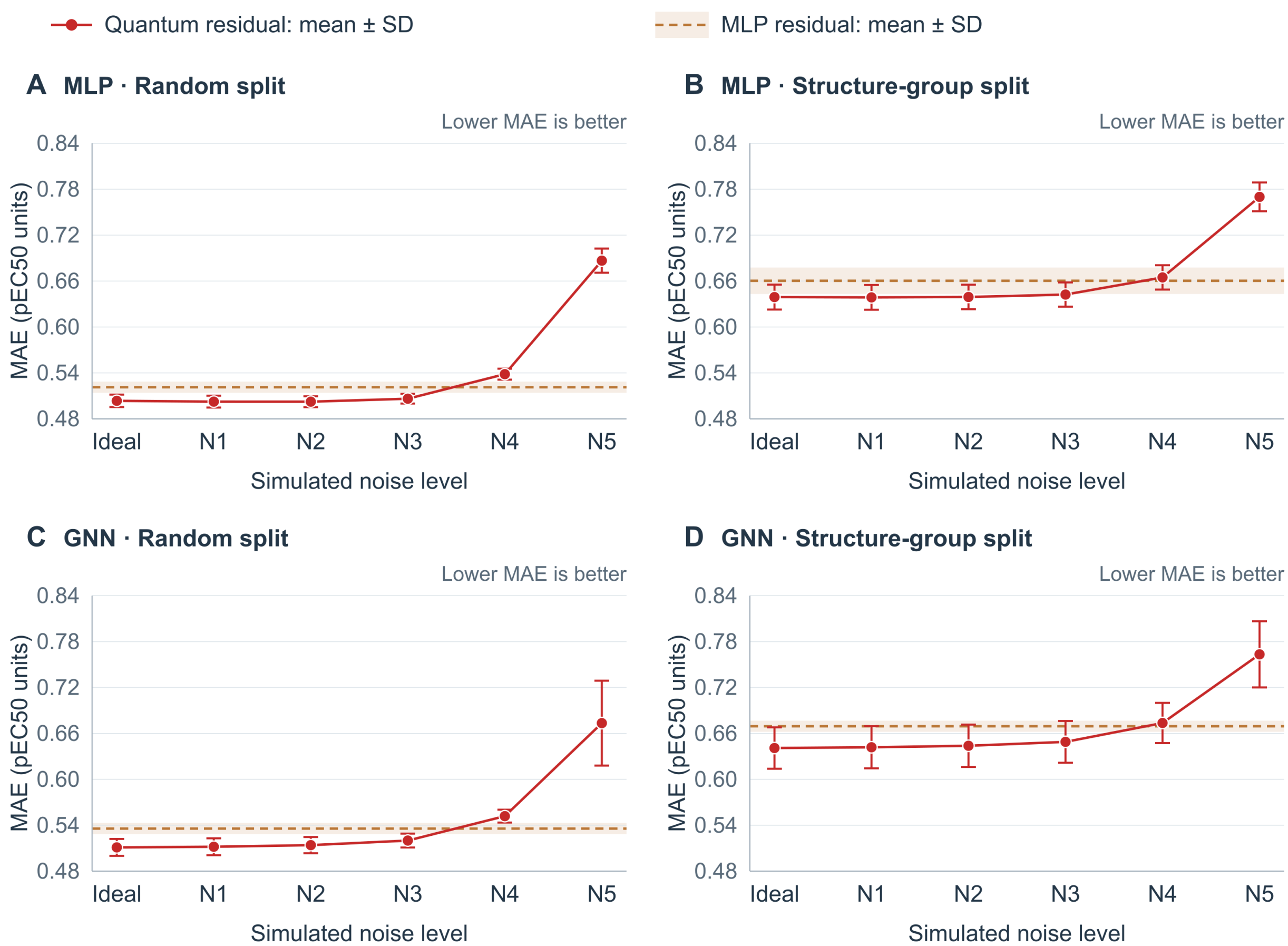

Figure S1: GLP-1R prediction error under simulated inference noise. (A,B) MLP encoder under random and structure-group splits, respectively; (C,D) GNN encoder under the corresponding splits. Red points and error bars show quantum-residual mean MAE and sample standard deviation across training seeds. The orange dashed line and shaded band show the corresponding classical MLP-residual mean and sample standard deviation. Ideal denotes noise-free inference; N1–N5 are the joint noise settings in Table S7a, displayed as ordered categories. Lines connect evaluated settings. Model parameters remain fixed, and simulated quantum noise is applied only to the quantum-residual models.

Table S7b: MAE and paired changes under simulated noise (mean ± sample standard deviation).

| Encoder | Condition | MAE | Noise-induced MAE change | MAE improvement over base |
|---|---|---|---|---|
| **Random split** | | | | |
| MLP | Ideal | 0.504 ± 0.008 | — | Table S6a |
| MLP | N1 | 0.502 ± 0.008 | -0.001 ± 0.001 | 0.025 ± 0.011 |
| MLP | N2 | 0.502 ± 0.007 | -0.001 ± 0.002 | 0.025 ± 0.010 |
| MLP | N3 | 0.506 ± 0.006 | 0.003 ± 0.004 | 0.021 ± 0.009 |
| MLP | N4 | 0.538 ± 0.007 | 0.035 ± 0.008 | -0.011 ± 0.009 |
| MLP | N5 | 0.687 ± 0.016 | 0.183 ± 0.017 | -0.159 ± 0.018 |
| GNN | Ideal | 0.511 ± 0.011 | — | Table S6a |
| GNN | N1 | 0.512 ± 0.011 | 0.001 ± 0.000 | 0.038 ± 0.013 |
| GNN | N2 | 0.514 ± 0.011 | 0.003 ± 0.001 | 0.036 ± 0.013 |
| GNN | N3 | 0.520 ± 0.009 | 0.009 ± 0.003 | 0.030 ± 0.011 |
| GNN | N4 | 0.552 ± 0.008 | 0.041 ± 0.016 | -0.002 ± 0.011 |
| GNN | N5 | 0.673 ± 0.055 | 0.162 ± 0.065 | -0.123 ± 0.056 |
| **Structure-group split** | | | | |
| MLP | Ideal | 0.639 ± 0.016 | — | Table S6a |
| MLP | N1 | 0.639 ± 0.016 | 0.000 ± 0.000 | 0.038 ± 0.009 |
| MLP | N2 | 0.639 ± 0.016 | 0.000 ± 0.001 | 0.037 ± 0.009 |
| MLP | N3 | 0.642 ± 0.016 | 0.003 ± 0.002 | 0.034 ± 0.008 |
| MLP | N4 | 0.665 ± 0.016 | 0.026 ± 0.004 | 0.012 ± 0.008 |
| MLP | N5 | 0.770 ± 0.019 | 0.131 ± 0.008 | -0.093 ± 0.010 |
| GNN | Ideal | 0.641 ± 0.027 | — | Table S6a |
| GNN | N1 | 0.642 ± 0.027 | 0.001 ± 0.001 | 0.038 ± 0.031 |
| GNN | N2 | 0.644 ± 0.028 | 0.003 ± 0.002 | 0.036 ± 0.031 |
| GNN | N3 | 0.649 ± 0.027 | 0.008 ± 0.004 | 0.031 ± 0.030 |
| GNN | N4 | 0.674 ± 0.026 | 0.033 ± 0.014 | 0.006 ± 0.026 |
| GNN | N5 | 0.763 ± 0.043 | 0.122 ± 0.050 | -0.084 ± 0.035 |

Under N1–N3, quantum-residual models retained lower mean MAE than their corresponding base models with both encoders and split schemes. At N4, this improvement persisted under structure-group splitting, whereas mean MAE slightly exceeded the base-model values under random splitting. At N5, mean MAE exceeded the base-model values in all four settings. Relative to the classical MLP-residual reference, mean MAE was lower under N1–N3 and higher under N4–N5 in all four settings (Figure S1). Small MAE reductions relative to ideal inference occurred in some low-noise settings.

Paired differences were calculated before rounding. Unrounded values are available in the “Noise Results” worksheet of Supplementary Data S1.

## S8 Data and Code Availability

The repository at `https://github.com/SpinQTech/Quantum-Residual-Potency` provides processed structures and labels, fixed splits, molecular fingerprints, and code for training, inference, and evaluation, including simulated noise. Out-of-fold predictions cover all reported models and seeds 42–46; Chemprop tuning records are also included. Pretrained weights cover all five folds at seed 42 for the three targets and both split schemes: MLP/GNN base and residual models, and Transformer base, MLP-residual, and quantum-residual models.

Supplementary Data S1 contains five data worksheets with unrounded values: “Model Summary” (aggregate metrics), “Model by Seed” (per-seed metrics), “Paired Comparisons” (per-seed MAE differences and bootstrap intervals), “Subgroup Results” (subgroup metrics), and “Noise Results” (noise analyses).

The code is released under the MIT License. ChEMBL-derived data retain source identifiers, with terms of use documented in the repository.